\documentclass[journal,lettersize]{IEEEtran}
\usepackage{amsmath,amsfonts}
\usepackage{algorithmic}
\usepackage{array}
\usepackage{subfig}
\usepackage{textcomp}
\usepackage{stfloats}
\usepackage{url}
\usepackage{verbatim}
\usepackage{graphicx}
\usepackage{color}
\def\BibTeX{{\rm B\kern-.05em{\sc i\kern-.025em b}\kern-.08em
    T\kern-.1667em\lower.7ex\hbox{E}\kern-.125emX}}
\usepackage{balance}
\begin{document}
\title{Semantic Communications with Variable-Length Coding for Extended Reality}
\author{Bowen Zhang~\IEEEmembership{Graduated Student Member,~IEEE}, Zhijin Qin~\IEEEmembership{Member,~IEEE}, and Geoffrey Ye Li~\IEEEmembership{Fellow,~IEEE}}

%\markboth{Journal of \LaTeX\ Class Files,~Vol.~18, No.~9, September~2020}%
%{How to Use the IEEEtran \LaTeX \ Templates}

\maketitle

\begin{abstract}
Wireless extended reality (XR) has attracted wide attentions as a promising technology to improve users' mobility and quality of experience. However, the ultra-high data rate requirement of wireless XR has hindered its development for many years. To overcome this challenge, we develop a semantic communication framework, where semantically-unimportant information is highly-compressed or discarded in semantic coders, significantly improving the transmission efficiency. Besides, considering the fact that some source content may have less amount of semantic information or have higher tolerance to channel noise, we propose a universal variable-length semantic-channel coding method. In particular, we first use a rate allocation network to estimate the best code length for semantic information and then adjust the coding process accordingly. By adopting some proxy functions, the whole framework is trained in an end-to-end manner. Numerical results show that our semantic system significantly outperforms traditional transmission methods and the proposed variable-length coding scheme is superior to the fixed-length coding methods. 
\end{abstract}

%\begin{IEEEkeywords}
%Class, IEEEtran, \LaTeX, paper, style, template, typesetting.
%\end{IEEEkeywords}

\section{Introduction}
\IEEEPARstart{E}{xtended} reality (XR), including augmented reality (AR), mixed reality(MR), and virtual reality (VR), has shown great potential in immersive gaming, teleconferencing, and remote education, and is changing the way that humans interact with the computer-simulated environments. In XR, the users need to wear head-mounted devices (HMDs) to perceive the virtual contents. These XR HMDs usually have strict restrictions on weight, energy consumption, and heat dissipation, especially when the users have to wear them for a long time. Therefore, XR offloading, as a way to offload storage and computational costive tasks of XR to remote GPU-accelerated servers, has been widely considered. 

Recently, wireless XR has been proposed to increase users' mobility and quality of experience (QoE) \cite{akyildiz2022wireless,morin2022toward}. However, XR offloading through wireless networks is a challenging task. In some latency-critical XR applications, high-resolution images or three-dimensional (3D) data need to be transmitted in several milliseconds, resulting in the data rate requirements as high as tens of Gbps \cite{zhang2022semantic}, which is far above the achievable capacity of the existing
wireless networks with peak from 0.1 Gbps to 2.0 Gbps. Therefore, a new communication framework for supporting wireless XR is more than desired.

To overcome the ultra-high data rate requirement in wireless XR, one promising solution is to exploit semantic communications to significantly reduce the data rate requirements~\cite{qin2021semantic}. In the existing communication systems, the source data is transmitted in the form of bit streams and the communication system is optimized to minimize the bit-error rate. However, how the semantic meaning of the source data will be used to conduct a specific task is not considered. These semantic-unaware or task-unaware transmission method usually requires high data transmission rate or wide bandwidth. Sometimes, even if there are some bit errors during transmission, humans will not misunderstand the semantic meaning of the recovered data or task accuracy will not decrease significantly. This phenomenon inspires us to develop semantic communications for wireless XR.   
 
Semantic communications transmits the semantic meaning in the source data, which is usually the content contributing more to the humans' understanding of the source data or task accuracy. Earlier attempt on semantic communications has developed the joint source and channel coding method (JSCC) for images \cite{bourtsoulatze2019deep} and texts \cite{farsad2018deep}, which first takes source coding into consideration and optimizes the system to minimize word/pixel-level errors rather than bit error. Semantic communications have first brought into wide attentions in~\cite{xie2021deep}, where BERT, a pre-trained natural language processing (NLP) model, is used to measure the sentence similarity in a semantic-aware wireless text transmission framework. The similar idea is then extended to the transmission of other data modalities, such as speech \cite{weng2021semantic}, images \cite{hu2022robust, dai2022nonlinear, huang2022toward}, and videos \cite{jiang2022wireless}. In addition to these reconstruction-oriented transmission works, task-oriented semantic communications have also been developed, which are more effective in reducing data traffic as only task-related semantic information is transmitted. Some representative works consider wireless image retrieval \cite{jankowski2020wireless,hu2022robust}, machine translation \cite{xie2022task}, visual question answering \cite{xie2021task} and so on. Despite the fast development of semantic communications, a semantic communication framework dedicated to wireless XR is still under-developed. 

Besides, most existing semantic communication systems adopt fixed code length for transmitting different source inputs~\cite{xie2021deep,bourtsoulatze2019deep}, ignoring the fact that different source inputs may contain different amounts of semantic information. A straightforward example is that the amount of semantic information contained in a picture of a coffee shop changes with the number of customers inside. Therefore, a shorter code should be used for source data with less semantic information. Also, different source inputs may have different anti-noise capabilities. For example, if we calculate the first derivative of the task performance index to the semantic information extracted from source data, a higher value usually means this information is more sensitive to channel noise and a longer code \footnote{when the number of information symbols is fixed, a longer code here means adding more parity symbols} should be used to conquer channel noise. Thus, variable-length coding scheme shall be introduced.

In this paper, we design a universal variable-length semantic-channel coding method for different semantic-aware transmission tasks. In particular, we first use a rate-allocation network to analyze the amount of semantic information and anti-noise capability extracted from the source data, and then generate a rate-allocation index/map. The rate-allocation index/map will then be used to guide the semantic-channel coding network and achieve rate adaption at the transmitter side. We also design a training loss function to realize the explicit trade-off between code length and transmission performance. By adopting some proxy functions, the whole system is trained in an end-to-end manner. 

Some prior works have considered exploiting variable-length coding scheme to semantic communications, but their methods differ from ours in certain ways. For instance, J. Dai \textit{et} \textit{al}. \cite{dai2022nonlinear} estimate the entropy of the semantic information and assign the code length proportionally to the estimated entropy. D. Huang \textit{et} \textit{al}. \cite{huang2022toward} divide the semantic features into different classes and adjust the quantization level for each class. However, these works have used strong hand-crafted assumptions on the coding rate allocation scheme while the optimal coding scheme may change from one transmission task to another. Therefore, we propose to learn the rate allocation scheme in an end-to-end manner. The most relevant work to ours is the one proposed in Q. Hu \textit{et} \textit{al}. \cite{hu2022robust}, where an importance weight is learned for each semantic feature. Different from \cite{hu2022robust}, we provide an explicit trade-off between code length and performance through the training loss.   
%as the weights for activated features must be larger than the threshold, while at the same time, have a summation smaller than $1-$ threshold,

%In addition, another major challenge of semantic communications is to find a training method to keep the semantic meaning of transmitted data during transmission process, and calculate the semantic loss through an objective quantitative index. Recent work on deep image compression \cite{mentzer2020high} has shown that using generative adversarial network (GAN) loss during training can help improve the perceptual quality of images, and indexes that calculate the distance between the distribution of source image and the distribution of the reconstructions image coincide with humans' subjective evaluation of image qualities, such as Fréchet Inception Distance (FID) \cite{heusel2017gans} and Kernel Inception Distance (KID) \cite{binkowski2018demystifying}. And some existing works have adopted similar idea to design and evaluate the semantic image transmission systems \cite{huang2022toward}. However, these works are restricted to image-type data only. Following the ideas of \cite{mentzer2020high}, we propose a general semantic communication framework that can be extended to other data type, such as 3D point cloud. 

Our contribution can be summarized as follows:
\begin{enumerate}
\item{We design a general semantic communication framework for supporting wireless XR offloading. We also identify the key transmission tasks under the developed framework, and provide task-aware network architectures for semantic coding modules.} 
\item{We design a universal variable-length semantic and channel coding module that can be used in different semantic communication systems. The rate-allocation scheme is learned in an end-to-end manner by introducing some proxy functions.}
\item{Experiments on both human mesh reconstruction and image transmission tasks demonstrate the superiority of the proposed semantic communication framework for wireless XR over the traditional communication systems, and the variable-length coding scheme over fix-length coding scheme.}
\end{enumerate}

\section{Semantic coding in wireless XR offloading}
\label{sec:overall}
\begin{figure*}[t]
\centering
\includegraphics[scale=0.45]{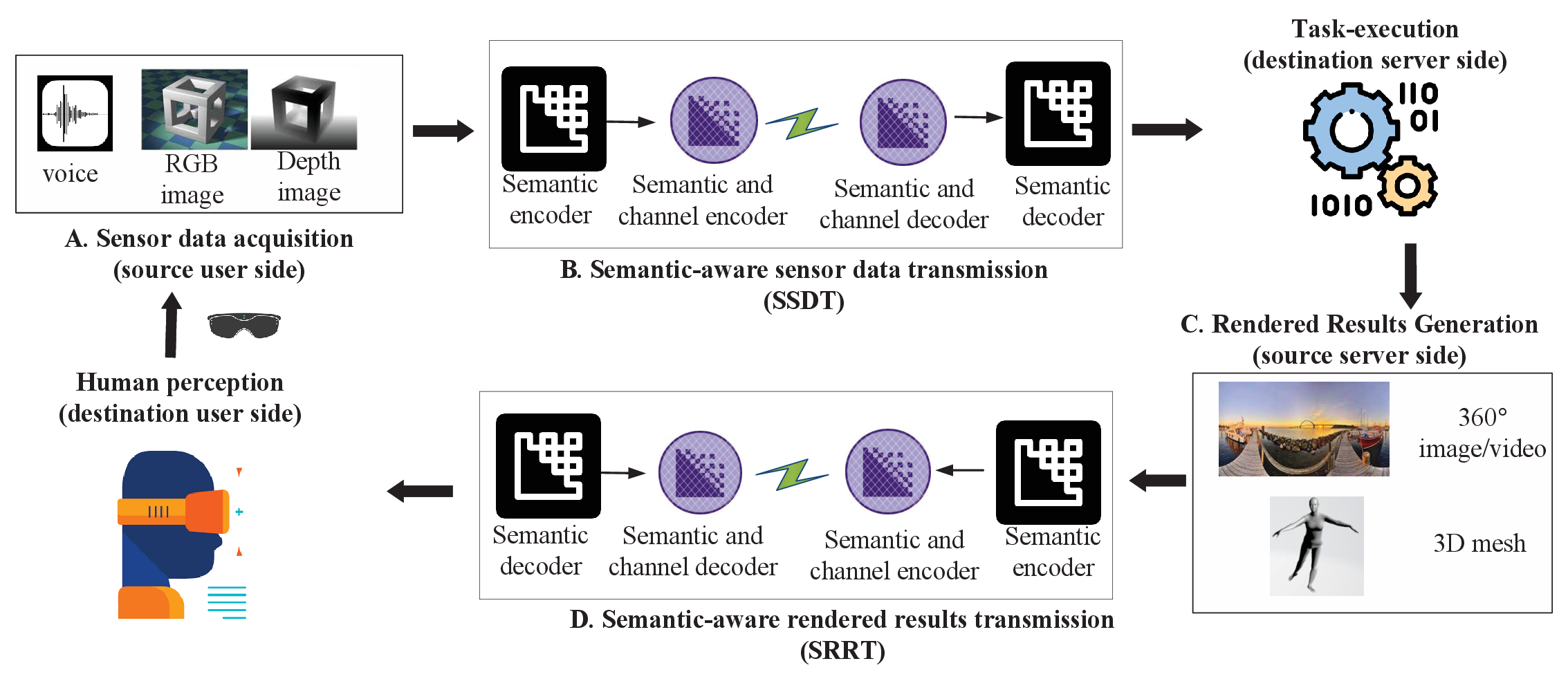}
\caption{The overall architecture of semantic communications for extended reality.}
\label{fig:overall}
\end{figure*}
In this section, we will first introduce the overall semantic communication framework for wireless XR offloading. After that, we will introduce the semantic coding modules deployed for the uplink and downlink of wireless XR offloading, respectively. 
\subsection{Overall architecture}
As shown in Fig.\ref{fig:overall}, semantic-aware wireless XR offloading usually consists of the following steps: sensor data acquisition, semantic-aware sensor data transmission, rendered results generation, and semantic-aware rendered results transmission. To facilitate the understanding of semantic-aware transmission modules, we first provide a brief description for the overall architecture.

\subsubsection{Sensor data acquisition}
In many XR applications, the virtual contents are generated according to the actions or environmental surroundings of XR users. For example, virtual object insertion requires users' environmental information; VR gaming servers will generate game scenes according to users' actions. To map objects, scenes, and human actions from the real environments to the virtual world, some sensors are required, such as microphones, RGB cameras, depth sensors, and so on. These sensors are usually deployed on XR HMDs, or integrated into some external devices, like haptic gloves and omnidirectional treadmills. Depending on the sensor type, the sensed data can be of various data formats. After collecting the sensed raw data, data aggregation,
coding, and compression will be performed at the user side. In XR offloading, these sensed data will be transmitted to remote servers to process. Here, we define the users who send sensed data to remote servers as source users, and the remote servers who receive sensed data from these source users as destination servers.

%Recently, with the development of computer vision technology, estimating environmental information, object's 3D architecture, and human actions from a single RGB or RGB-depth image has attracted wide attentions. These methods can free XR users from using various external equipment and reduce the sensing costs. Therefore, image-type sensed data will become the mainstream in the future wireless XR offloading.

\subsubsection{Semantic-aware sensor data transmission}
In traditional communication systems, the sensed data is first transmitted to the remote destination servers, and then used for different computing tasks, such as light estimation, 3D object reconstruction, and object insertion. Considering the fact that task-related semantic information only occupies a small proportion of the overall information contained in the sensed data, transmitting the whole sensed data wastes wireless resources. Therefore, we design a semantic-aware sensor data transmission (SSDT) module.

In our SSDT module, a semantic encoder deployed at the XR source user side, is responsible for extracting task-related semantic information from the sensed data. By discarding semantically-unimportant information, the data traffic for transmitting sensor data can be reduced. At the same time, a semantic decoder deployed at the XR destination server side will use these semantic information directly for task execution. The whole system is then optimized for transmitting these task-related information under the guidance of task accuracy. More details will be provided in Section \ref{sec:SSDT}. 

\subsubsection{Rendered results generation}
Based on the semantic decoding results from the SSDT module, a series of rendering operations will be conducted at the XR server side. Specifically, for VR applications, the status of the pre-stored 3D models of objects or scenes will first be updated. This includes adding new 3D object models to the virtual environments, changing the orientations and actions of the 3D object models, and refurbishing the outlooks of the 3D object models. Afterwards, photo-realistic viewpoint-dependent perspective images or 360° panoramic images will be rendered for XR users to perceive these 3D scenes, depending on the XR users' current field of views (FoVs). As for AR/MR applications, in addition to changing the orientations or appearances of the 3D object models, light or shadow effects will also be added to these models according to the environmental changes near XR users. After rendering, these 3D object models will be transmitted for targeted users to perceive. 

In some XR applications, the users who receive rendered results from XR servers can be the same ones who send sensor data. For example, AR game servers will require users' to send their environment information and integrate virtual game elements into the physical environments of the same users. While in other applications like VR conferencing and multiplayer VR games, the sensed data transmitted by one user will be used to render the virtual contents designed for other users. Without loss of generality, we call the servers who send rendered results as source servers while the users who receive rendered results as destination users.  

\subsubsection{Semantic-aware rendered results transmission} 
In traditional communication systems, the rendered images or 3D object models are transmitted to destination users pixel-by-pixel or point-by-point. However, as discussed in the previous works \cite{blau2018perception,mentzer2020high}, simply minimizing the per-pixel error between the original images and the reconstructed images does not necessarily result in high perceptual quality. To improve users' QoE, perceptual quality is considered in our SRRT module. Besides, due to the round-trip latency in XR offloading, XR users' FoVs at the time of receiving these rendered results may be different from the FoVs in the sensor data acquisition stage. In this case, these rendered results shall be transmitted in a way that is friendly to local view-synthesis process. To address these issues, we design a semantic-aware rendered results transmission (SRRT) module.

In our SRRT module, a semantic encoder is deployed at the XR source server side to extract perceptual-friendly or view-synthesis-friendly semantic information from the rendered results. Similarly, a semantic decoder deployed at destination user side will generate or synthesis viewpoint-dependent images and 3D object models using the received semantic information. The whole system is then optimized to maximize perceptual experience, as we will discuss in more details in Section \ref{sec:SRRT}.

\subsection{Uplink SSDT module}
\label{sec:SSDT}
In XR applications, the sensed data is used for task execution. 
Sometimes, both the sensed data and the task outputs can be with a large size and unsuitable for transmission. To reduce the data traffic, we develop the SSDT module for XR offloading. In the proposed SSDT module, task-related semantic information, which lies in a hidden space smaller than both the sensed data space and task outputs space, is extracted from the sensed data and used for task execution through a semantic encoder and decoder. Different from the traditional coding modules, the semantic information extracted by semantic coding modules has clear semantic meanings, which is learnt either by proving extra labels or under the guidance of a model-based decoder. 
%After extracting these semantic information, joint semantic and channel coding modules are used to map these semantic information into modulated symbols that are suitable for transmission. The details of joint semantic and channel coding module will be given in Sec. 
In the subsequent discussion, We take some classic computing tasks in XR offloading as examples and introduce the corresponding semantic coding modules.
\subsubsection{Light estimation}
Light estimation aims to recover a high-dynamic-range (HDR) panoramic illumination map from a single image with limited FoV, which can later be used in many AR/MR applications for realistic virtual object relighting.
%An accurate light estimation can ensure the inserted objects have clear and correct shade and shadows.
In light estimation tasks, the inputs are high-resolution (HR) perspective RGB images captured by HMD cameras and the outputs are HDR panoramic illumination map. XR source users can choose to either send the HR images to XR destination servers or process HR images locally and send estimated illumination map to destination XR servers. However, due to the large data volume of HR images and panoramic images, both ways are challenging to wireless networks. To address this issue, we refer to the recent light estimation method EMLight \cite{zhan2021emlight} and design the semantic coding modules accordingly. 

In particular, a semantic encoder, composed of a regression neural network (DenseNet-121) and a feature encoder, is deployed at the XR user side. The regression network extracts high-level semantic features from input RGB images and regresses these features into $N(=128)$ 3D anchor points representing light distribution, a 3D light intensity value, and a 3D ambient term. Simultaneously, the feature encoder encodes the input images and generates a $L$ ($\leq 2048$)-feature vector representing source users' surrounding environments. At the destination server side, a spherical convolution network, served as the semantic decoder, takes the received semantic information as inputs and synthesizes the illumination map via conditional image synthesis process. Therefore, the total number of semantic information after the semantic coding modules is $3(N+2)+L$, which is significantly smaller than the original HR images and expected HDR panoramic illumination output. %These semantic information can be further compressed and channel-coded in joint semantic and coding modules before final transmission.  
\subsubsection{3D mesh recovery}
3D mesh recovery of human hand, face, and body shape from a single RGB image has wide applications in VR conferencing and gaming. In this field, a popular direction is to train a regression network for fitting parameters of parametric 3D hand/face/body models, such as hand model with articulated and non-rigid deformation (MANO) for hands \cite{zhang2019end}, 3D morphable model (3DMM) for faces \cite{tewari2017mofa}, and skinned multi-Person linear model (SMPL) for body shape \cite{kanazawa2018end}, and use estimated parameters for 3D mesh reconstruction. These methods are quite suitable for semantic communications, as we only need to transmit these estimated parameters among the source users and destination servers, and the latter can use received parameters for 3D mesh recovery. Compared with the original RGB images or generated 3D meshes, these parameters are quite small. For example, 76 semantic features are required for 3D hand mesh recovery, among which $3$ features are for camera orientation, $10$ features for hand shape, and $21$ 3D keypoints for hand pose. In this case, the semantic encoder is the regression network while the semantic decoder is the MANO model. Similar process can be applied to face mesh ($257$ semantic features) and human body and shape mesh ($85$ semantic features). A detailed transmission paradigm for human mesh recovery will be given in Section \ref{sec:human mesh}.

\subsubsection{3D scene reconstruction}
3D scene reconstruction from a sequence of RGB images is widely used in AR applications for virtual object placement. In 3D scene reconstruction task, the 3D scene will be represented by a 4D truncated signed distance function (TSDF) volume with color values. Both the image sequences and TSDF volume are with large sizes. Recent studies on sparse TSDF representation \cite{sun2021neuralrecon} and TSDF completion \cite{dai2018scancomplete} inspire us to transmit a sparse representation of TSDF for data traffic reduction. In our SSDT module, the semantic encoder at the source user side first finds a sparse TSDF volume representation from image sequences using the similar process shown in \cite{sun2021neuralrecon}, where only voxels representing surfaces have non-zero values. And then, a downsampling operation is applied to the sparse TSDF volume, as an incomplete 3D scan of these 3D surfaces. At the destination server side, a ScanComplete network \cite{dai2018scancomplete} is used as the semantic decoder to reconstruct the original sparse TSDF volume, followed by a global TSDF volume replacement defined in \cite{sun2021neuralrecon} for final TSDF volume reconstruction.

\subsection{Downlink SRRT module}
\label{sec:SRRT}
After rendering, the downlink SRRT module will transmit rendered results to destination users in a semantic-aware manner. In particular, semantic information that is important for users' visual QoE will consume more channel resources during the transmission process so that users' visual QoE can increase. Here, we consider two factors affecting users' visual QoE, \textit{i}.\textit{e}., perceptual quality, and FoV mismatch. We will design semantic coding modules for these factors as follows.  
\subsubsection{Perceptual-friendly XR contents transmission} To improve the perceptual quality of reconstructed XR images, we use three perceptual-friendly semantic losses to enhance the semantic information extraction and reasoning process in semantic coding modules: learned perceptual image patch similarity (LPIPS) \cite{zhang2018unreasonable} loss, generative adversarial network (GAN) loss \cite{mentzer2020high}, and salience-weighted loss \cite{wang2023robust}.  
LPIPS loss measures the distance of the original images and the recovered images in the feature space of a deep neural network originally
trained for image classification. Minimizing the discrepancy in feature space has been regarded as a way to improve perceptual quality in earlier works \cite{zhang2018unreasonable}. GAN uses a learned discriminator to detect the artifacts in the recovered images and adopts adversarial training to encourage the semantic coding modules to generate artifact-free contents. Salience-weighted loss imposes a higher penalty to the pixels belong to salient objects so that the salient objects in images can be constructed better. As users first notice the salient objects in images, salience-weighted loss can help improve visual QoE. Under the guidance of these semantic-aware losses, semantic information affecting human's visual perception process can be extracted more effectively. 

The implementation details of LPIPS and GAN losses will be provided in Section \ref{sec:examples}. To implement salience-weighted loss, a salience detection network is required \cite{wang2015deep}, whose outputs will be used to design the weighted loss. To boost the performance of salience-weighted loss, a spatially-varying coding scheme shall be implemented, so that pixels do not belong to any salient object can be assigned a short code, as they are more likely to be ignored by users. In Section \ref{spatial}, we will develop a spatially-varying coding method.

\subsubsection{View-Synthesis for XR contents} Due to fast changes of users' viewing directions and standing points, XR users' current FoVs may be different from the FoVs in the sensor data acquisition process, leading to the FoV mismatch between rendered contents and ideal contents. Despite the FoV mismatch caused by viewing direction changes can be solved by transmitting panoramic images, FoV mismatch resulting from standing points changes requires further signal processing techniques. In the SRRT module, we adopt existing view-synthesis technologies \cite{wiles2020synsin} to address this problem and design the semantic coding modules correspondingly. Specifically, a semantic encoder consists of a feature predictor and a depth regressor is deployed at the source server side. The extracted semantic features and depth map are then transmitted to the destination users. Based on users' current FoV, a neural render and a refinement module are used as a semantic decoder to synthesize the targeted views. Transmitting these semantic features with clear semantic meanings and functionalities enables the fast viewpoint adaptation at the destination user side. 

\section{Variable-length semantic and channel coding}
\label{sec:VLCC}
\begin{figure*}[t]
\centering
\includegraphics[scale=0.48]{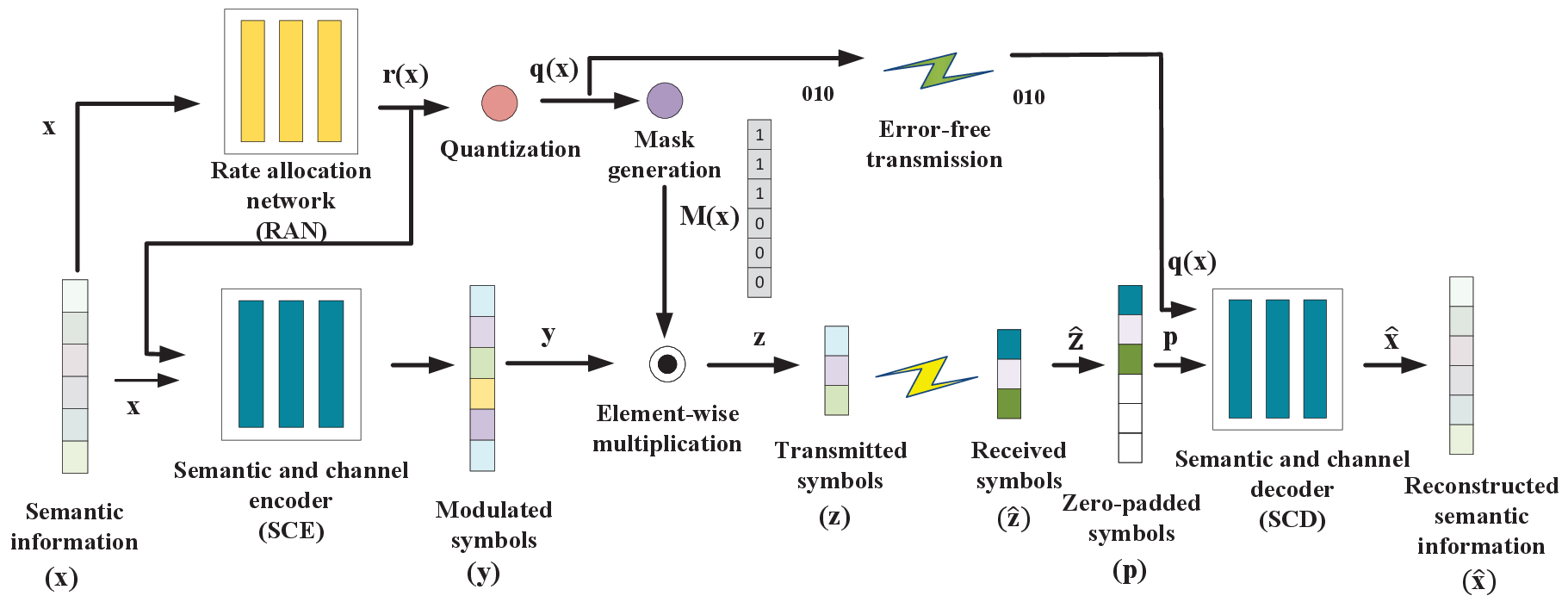}
\caption{The variable-length semantic-channel coding method for 1D semantic information.}
\label{fig:adaptive}
\end{figure*}
In this section, we first introduce the proposed variable-length semantic-channel coding method (VL-SCC) for 1D semantic information. And then, we will extend it to 2D/3D semantic information with spatially-varying coding scheme.

\subsection{VL-SCC for 1D semantic information}
\label{sec:1DVL}
\subsubsection{Overall architecture}
The overall architecture of the proposed VL-SCC is shown in Fig. \ref{fig:adaptive}. Given a 1D semantic information $x$, we first apply a rate allocation network (RAN) to $x$ to analyze its amount of semantic information and anti-noise capability and output a rate allocation index, $r(x) \in (0,1)$. Then, $r(x)$ is used to guide the coding process and generates a $0$-$1$ mask for rate control. 

In particular, a semantic and channel encoder (SCE), another neural network, takes both $x$ and $r(x)$ as inputs and generates $N$ modulated symbols $y\in \mathcal{R}^{N}$. Under the guidance of $r(x)$, $y$ will be generated in a specific way, \textit{i}.\textit{e}., the most important information symbols and parity symbols for $x$ transmission are selected and put in the first $Nr(x)$ symbols of $y$ while the other symbols are generated in an importance-descending way.

At the same time, a $0$-$1$ mask, $M(x)\in \mathcal{R}^{N}$, will be generated according to the rate index $r(x)$. Specifically, we first apply a uniform quantizer to $r(x)$, and quantify $r(x)$ into $L$ discreet values. Denote the quantization result as $q(x)$. And then, we generate $M(x)$ according to the value of $q(x)$ in a way that $1$ values are all generated ahead of $0$ values. The details of the quantizer and mask generation operation will be given hereafter. Once generated, $M(x)$ will be element-wisely multiplied with $y$ to get transmitted symbols, $z$, and the symbols with $0$ values in the mask can be safely discarded before transmission process. We now finish the coding rate adaption at the encoder side. 

After encoding, the shortened transmitted symbols, $z$, will be transmitted to the decoder through noisy wireless channels. Simultaneously, $q(x)$, indicating the adopted variable-length coding scheme at the encoder side, will also be transmitted to the decoder. Different from $z$, $q(x)$ will be transmitted in an error-free manner. As the quantization level, $L$, is chosen as a small value, $q(x)$ can be represented by several bits and can be easily transmitted without error. 

After receiving the noise-corrupted symbols $\hat{z}$, we first apply zero-padding to them to ensure that they have a length of $N$. After this, these zero-padded symbols, $p$, and $q(x)$ are concatenated and fed into a semantic and channel decoder (SCD), which has symmetric architecture with the SCE. Finally, semantic information is reconstructed at the decoder side, which is annotated as $\hat{x}$. 

As we can see from Fig. \ref{fig:adaptive}, the quantizer and mask generation process restrict the end-to-end training property of the whole architecture. Also, a training loss should be designed to train the SCE, SCD, and rate allocation network. In the following, we will address these issues sequentially.

\subsubsection{Quantizer} 
We first solve the gradient problem of quantizer. In our network, the quantization process is defined as as follows,
\begin{equation}
q(x)=l, \text{if} \, 
\label{equ:1}
\end{equation}
$\frac{l-0.5}{L-1}\leq r(x) \leq \frac{l+0.5}{L-1}$, for $l=0,1, \dots, L-1$. As mentioned above, $r(x)\in (0,1)$. Therefore, $q(x)$ may take one of $L$ different quantity values, \textit{i}. \textit{e}. $0, 1, \dots, L-1$. After using this quantization formulation, the gradient is zero almost everywhere. To address this issue, we use a straight-through estimator of the gradient \cite{courbariaux2016binarized} and define the gradient in the back-propagation process as,
\begin{equation}
q'_{x}=L-1.
\label{equ:2}
\end{equation}
In this way, the gradient can be back-propagated from $q(x)$ to $r(x)$.
\subsubsection{Mask generation} Next, we will introduce the design of mask generation operation. In the forward-propagation process, we first initialize an all-zero vector $M(x)$. Then, we define the value of each element in $M(x)$ as follows,
\begin{equation}
\begin{split}
 m_{i}= \left \{
\begin{array}{ll}
1,     & i < \frac{N}{L-1} q(x),\\
0,     & \text{else},\\
\end{array}
\right.
\end{split}
\label{equ:3}
\end{equation}
for $i=0,\dots, N-1$, where $m_{i}$ is the $i$-th element in $M(x)$. As shown in Eq. (\ref{equ:3}), the value of $m_{i}$ is decided by a step function of $i$. Given a specific $q(x)$ value, all elements of $M(x)$, whose corresponding locations in $M(x)$ are smaller than $\frac{N}{L-1} q(x)$, have $1$ values while the rests have $0$ values, In this way, we can ensure all $1$ values are generated before $0$ values in the final $0$-$1$ mask. Also, when $q(x)$ has a larger value, $\frac{N}{L-1} q(x)$ will increase, making more elements in $M(x)$ taking $1$ values. 

Similar to the quantization operation, the gradient is zero everywhere for $\frac{dq}{dx}$. To address this issue, we first rewrite Eq. (\ref{equ:3}) as,
\begin{equation}
\begin{split}
 m_{i}= \left \{
\begin{array}{ll}
1,     & q(x) > \frac{L-1}{N} i,\\
0,     & \text{else},\\
\end{array}
\right.
\end{split}
\label{equ:4}
\end{equation}
for $i=0,\dots, N-1$, to make each $m_{i}$ as a step function of $q(x)$. Under this formulation, we can again adopt a straight-through estimator of the gradient. We define the gradient in the back-propagation process as, 
\begin{equation}
\begin{split}
 {m'_{i}}_{q}= \left \{
\begin{array}{ll}
1/3,     & \lceil \frac{L-1}{N} i\rceil-2<q(x)\leq \frac{L-1}{N} i\rceil+1,\\
0,     & \text{else},\\
\end{array}
\right.
\end{split}
\label{equ:4}
\end{equation}
where $\lceil \rceil$ is the ceil operation. Here, the gradient is defined as an interval of $3$ units to stabilize the training process. Under this formulation, only $m_{i}$, whose location in $m$ is close to $\frac{N}{L-1} q(x)$, will contribute gradients to $q(x)$.  

\subsubsection{Model learning} At last, we will define the training process for the whole network. Specifically, we want the encoded representation of semantic information to be compact when measured in symbols while at the same time
we want distortion $\mathcal{L}_{d}(x,\hat{x})$ to be small, where $\mathcal{L}_{d}$ is some measurement of the reconstruction error. Under this design goal, we can formulate the training loss as the well-known rate-distortion trade-off \cite{mentzer2020high,balle2016end,mentzer2018conditional}. Let $\mathcal{X}$ be a set of semantic information used for training and $x\in \mathcal{X}$ be a training example from the set, the training loss is defined as,
\begin{equation}
\mathcal{L}=\sum_{x\in \mathcal{X}}{\mathcal{L}_{D}(x,\hat{x})+\gamma \mathcal{L}_{r}(x)},
\label{equ:5}
\end{equation}
where $\mathcal{L}_{r}(x)$ represents the rate loss and $\gamma$ is an introduced trade-off parameter between distortion loss and rate loss. Increasing the value of $\gamma$ will penalize more on the coding length and reduce the average code length of $\mathcal{X}$ sets.

\textbf{Distortion loss:} Distortion loss is used to evaluate the
transmission quality of semantic information. In semantic communications, it is a combination of a data-fidelity term (for faithful reconstruction) and a semantic loss term (for human understanding). As its formulation is task-specific, we do not give its detailed mathematical expression here but leave it to Section \ref{sec:examples}, where some semantic-aware XR applications transmission task will be given.

\textbf{Rate loss:} 
As shown in Fig. \ref{fig:adaptive}, the number of transmitted symbols is completely decided by the value of $r(x)$. Therefore, $r(x)$ can be regarded as a continuous indicator of code length. We define the rate loss as $\mathcal{L}_{r}(x)=r(x)$.

By minimizing $\mathcal{L}_{D}(x,\hat{x})$, more elements in $M(x)$ will be driven to be $1$ values while by minimizing $\mathcal{L}_{r}(x)$, more elements in $M(x)$ will be $0$ values, achieving an explicit trade-off between the performance and the code length. Through Eq. (\ref{equ:5}), the best code for each $x$ can be learnt.
\begin{figure*}[ht]
\centering
\includegraphics[scale=0.75]{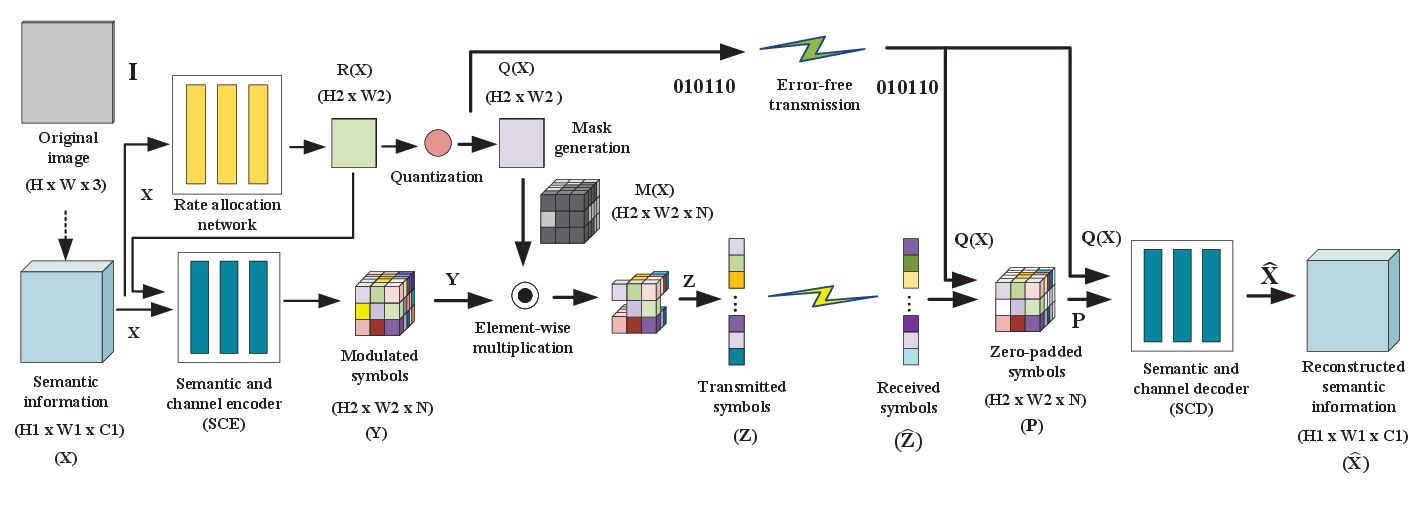}
\caption{The variable-length semantic-channel coding method for 2D/3D semantic information.}
\label{fig:adaptive_image}
\end{figure*}
\subsection{Spatially-variant VL-SCC}
\label{spatial}
In the previous subsection, the variable-length coding scheme is deigned for 1D semantic information. Here, we will extend it for 2D/3D semantic information by introducing spatially-variant coding. Parts of the design here are inspired by the early deep compression work in \cite{li2018learning}. Different from \cite{li2018learning}, we jointly consider source and channel coding and directly map semantic information into modulated symbols without bit transition.
\subsubsection{Motivation}
Given a 2D/3D semantic information, especially those extracted from images/videos, its information amount is spatially-variant\footnote{Given a 3D tensor of shape $(H \times W \times C)$, we define the first two dimensions as spatial dimension, and the last as channel dimension}. For example, the semantic information extracted from a sky region in images has less amount of information than those extracted from a human region. And if the average information amount of all regions is relatively-smaller, a shorter code should be used. This motivates us to design a spatially-variant coding.  

\subsubsection{Overall architecture}
The architecture of the proposed VL-SCC for 2D/3D semantic information is shown in Fig. \ref{fig:adaptive_image}. Given a semantic information $X \in \mathcal{R}^{H_{1}\times W_{1}\times C_{1}}$ extracted from an original image of shape $(H, W, 3)$, we first adopt a rate allocation network to analyze its semantic information amount and anti-noise capability in different spatial locations and generate a rate allocation map $R(X)\in \mathcal{R}^{H_{2}\times W_{2}}$, which indicates the final code length at each spatial location of modulated symbols. 

Different from the rate index in Section \ref{sec:1DVL}, $R(X)$ here is a matrix. The burden of transmitting its quantified version is proportional to its spatial size. To address this issue, down-sampling operation is adopted in the rate allocation network. 

Similarly, $R(X)$ is used in two ways. It is first concatenated with $X$ and fed into a SCE. The SCE will generate modulated symbols $Y$ of shape $(H_{2}, W_{2}, N)$. Under the guidance of $R(X)$, $Y$ will be generated to satisfy the following properties: 
\begin{enumerate}
    \item The most important information symbols and parity symbols at spatial location $(i,j)$ will be selected and put in the first $R(X)_{i,j}N$ symbols at that location\footnote{In semantic communications, different information symbols have different importance levels according to their contributions to the transmission task. Usually, each parity symbol is only used to protect several information symbols, like LDPC coding. In this case, parity symbols will also have importance differences according to the information symbols they are protecting. By training, the networks can learn to distinguish important information and parity symbols from others, and adjust their locations.};
    \item For location $(i,j)$ where the number of important symbols is smaller than $R(X)_{i,j}N$, it will accept important symbols from neighbouring locations $(m,n)$ where the number of important symbols is larger than $R(X)_{m,n}N$;
    \item The rest symbols are generated in an importance-descending way along the channel dimension. 
\end{enumerate}
In SCE, down-sampling operation is also used to ensure that the modulated symbols $Y$ have the same spatial size as $R(X)$.

Simultaneously, $R(X)$ will be quantified by Eq. (\ref{equ:1}) to get $Q(X)$. After that, a mask tensor, $M(X) \in \mathcal{R}^{H_{2}\times W_{2} \times N}$, is calculated, where each mask vector in the spatial dimension of $M(X)$ is generated using Eq. (\ref{equ:3}). Next, we will conduct an element-wise multiplication operation between $M(X)$ and $Y$, and discard symbols at locations in which the mask value is $0$. Finally, the rest symbols will be reshaped into a vector $Z$. 

After encoding, the symbol vector $Z$ will be transmitted to the decoder via noisy channels. Simultaneously, $Q(x)$ will be transmitted in an error-free way. The bit length of $Q(x)$ will be $H_{2}W_{2}\log_{2}(L)$. Considering the spatial size of the original image $HW$, the bit rate for this error-free transmission link will be $H_{2}W_{2}\log_{2}(L)/HW$ bits per pixel (bpp), which must be set as a small value to reduce the transmission costs for this error-free side link. In our methods, we set $L={\color{blue}{16}}$, $H_{2}=\frac{H}{16}$, and $W_{2}=\frac{W}{16}$, resulting in ${\color{blue}{0.0156}}$ bpp. Also, as this is an error-free link, entropy coding can be used to further decrease the bit rates of transmitting $Q(x)$. %We adopt Huffman coding and reduce the bit rates to $0.0209\sim 0.0232$\footnote{As we do not introduce an extra loss to explicitly reduce the entropy of $Q(x)$, the rate reduction brought by entropy coding is not significant. We will consider it in our future work.}. 

At the decoder side, $Q(x)$ will guide the process of reshaping the received symbol vector $\hat{Z}$ back to a 3D tensor of shape $(H_{2},W_{2},N)$. Zero-padding is applied in this process to pixels that are discarded at the encoder side. After getting zero-padded symbols $P$, a SCD will reconstruct the 2D/3D semantic information, $\hat{X}$. 

\textbf{Rate loss:} 
As shown in Fig. \ref{fig:adaptive_image}, the code length at each spatial location $(i,j)$ is decided by $R(x)_{i,j}$. Therefore, we define the rate loss as $\mathcal{L}_{r}(x)=\sum_{i}\sum_{i}R(x)_{i,j}$.

\section{Joint Semantic Coding and Semantic-Channel Coding}
\label{sec:examples}
In this sections, we will show how the semantic coding modules developed in Section \ref{sec:overall} and the VL-SCC module designed in Section \ref{sec:VLCC} are jointly optimized in semantic communications. Specifically, we will introduce the implementation details for human mesh recovery task in the uplink SSDT module and the perceptual-friendly wireless image transmission in the downlink SRRT module.
\subsection{Human mesh recovery}
\label{sec:human mesh}
\begin{figure}[t]
\centering
\includegraphics[scale=0.65]{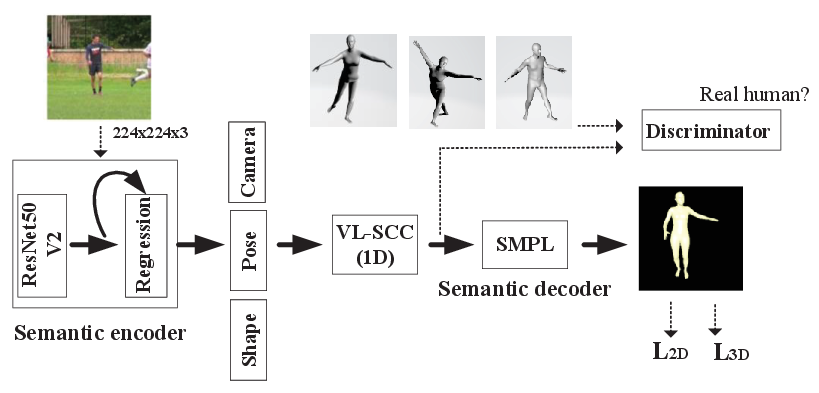}
\caption{Semantic communications for human mesh recovery in uplink SSDT.}
\label{fig:human_mesh}
\end{figure}
The semantic communication framework for human mesh recovery task in the uplink SSDT is shown in Fig. \ref{fig:human_mesh}. From the figure, a semantic encoder, consisted of a ResNet50 and a regressor, will take a RGB image as input and estimate task-related semantic information at XR source user side. The semantic information, $x \in \mathcal{R}^{85}$, parameterizes the camera parameters ($t \in \mathcal{R}^{3}$), human pose ($\alpha \in \mathcal{R}^{72}$, 3D rotation of 23 joints + 1 global rotation), and human body shape ($\beta \in \mathcal{R}^{10}$). These semantic information is then transmitted to and reconstructed at the destination server using the VL-SCC module introduced in Section \ref{sec:1DVL}. For the networks used in the VL-SCC module, we set them as fully-connected network with $1024$ hidden units ($6$ layer used for SCE and SCD, $4$ layer for rate allocation network). The VL-SCC module will output the reconstructed semantic information $\hat{x}=(\hat{t}, \hat{\alpha}$, $\hat{\beta})$. At last, a SMPL model is used to reconstruct the final 3D triangulated human mesh with $N = 6980$ vertices, $z\in \mathcal{R}^{N\times 3}$, from $\hat{x}$.

To jointly train the semantic coding module and the VL-SCC module, a training dataset and a training loss are required. In this task, we have access to three kinds of data sets: 
\begin{enumerate}
    \item A large RGB image dataset where all are annotated with ground truth 2D joints;
    \item A small dataset in which some have 3D joints annotations;
    \item A 3D human mesh dataset of varying shape and pose but without corresponding images.
\end{enumerate}

Based on the available datasets and earlier work \cite{kanazawa2018end}, we design the training loss correspondingly. The training loss is composed of three terms: a data fidelity term $\mathcal{L}_{l}$, a semantic loss $\mathcal{L}_{s}$, and a rate loss $\mathcal{L}_{r}$.

\textbf{Data fidelity term:} The data fidelity term is used to minimize the distance of 2D joints' locations and 3D joints' location between true data labels and those calculated from reconstructed 3D human mesh. Specifically, 3D joints' locations can be obtained by linear regression
from the reconstructed mesh vertices $z$; 2D joints' locations are calculated by projecting 3D joints locations to 2D using estimated camera parameters $t$. Thus, the data fidelity term can be represented as $\mathcal{L}_{f}=\mathcal{L}_{2D}+\mathcal{L}_{3D}$.

\textbf{Semantic loss:}
Semantic loss is designed for realistic-looking reconstruction of the 3D human mesh. This is because a low data fidelity term does not always mean that the reconstructed 3D human mesh has a good visual quality, especially when we only have partial observations on 2D and 3D joints, and have no access to a labelled human mesh dataset. To address this problem, a semantic loss based on GAN is developed. Specifically, we introduce a discriminator $D$, which is trained by minimizng
$E_{(\Theta \sim p_{T})}(D(\Theta-1)^2)+E_{(\hat{\Theta} \sim p_{R})}(D(\hat{\Theta})^2)$, where $\Theta=(\alpha,\beta)$; $\hat{\Theta}=(\hat{\alpha},\hat{\beta)}$; $p_{T}$ denotes the distribution of $\theta$ in true mesh dataset; $p_{R}$ denotes the distribution of estimated parameters. $D(\Theta)\in (0,1)$ denotes the output of the discriminator when $\Theta$ is input. With this training loss, the discriminator is able to distinguish the true parameters from the estimated ones. Simultaneously, the semantic loss, $\mathcal{L}_{Gan}$, is defined as $E_{(\hat{\Theta} \sim p_{R})}(D(\hat{\Theta}-1)^2)$, which will encourage the estimated parameters to have the same distribution with the true parameters.

\textbf{Rate loss:}
Rate loss is adopted to train the variable-length coding scheme in the VL-SCC module. It is formulated as $\mathcal{L}_{r}$ as before.

Finally, the total training loss can be represented as,
\begin{equation}
\mathcal{L}=\mathcal{L}_{f}+\lambda \mathcal{L}_{Gan}+ \gamma \mathcal{L}_{r},
\label{equ:6}
\end{equation}
where $\lambda$ is an introduced trade-off parameter between data fidelity term and semantic loss. A high value will penalize more on the distribution discrepancy.

\subsection{Perceptual-friendly wireless image transmission}
\begin{figure}[t]
\centering
\includegraphics[scale=0.7]{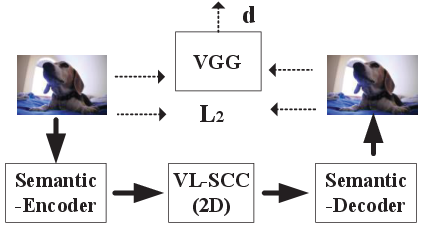}
\caption{Semantic communications for perceptual-friendly image transmission in downlink SRRT.}
\label{fig:image}
\end{figure}

\begin{table}[]
\centering
\caption{Networks for Wireless Image Transmission}
\label{sec:image}
\label{tab:my-table}
\begin{tabular}{|cccccc|}
\hline
\multicolumn{3}{|c|}{\textbf{Semantic Encoder}}                                             & \multicolumn{3}{c|}{\textbf{SCE}}                                   \\ \hline
\multicolumn{1}{|c|}{Type}   & \multicolumn{1}{c|}{Para.}   & \multicolumn{1}{c|}{Out\_c} & \multicolumn{1}{c|}{Type}   & \multicolumn{1}{c|}{Para.}   & Out\_c \\ \hline
\multicolumn{1}{|c|}{S-Conv} & \multicolumn{1}{c|}{9x9,S2} & \multicolumn{1}{c|}{256}    & \multicolumn{1}{c|}{S-Conv} & \multicolumn{1}{c|}{5x5,S2} & 512    \\
\multicolumn{1}{|c|}{PReLu}  & \multicolumn{1}{c|}{-}       & \multicolumn{1}{c|}{-}      & \multicolumn{1}{c|}{PReLu}  & \multicolumn{1}{c|}{-}       & -      \\
\multicolumn{1}{|c|}{S-Conv} & \multicolumn{1}{c|}{5x5,S2} & \multicolumn{1}{c|}{256}    & \multicolumn{1}{c|}{S-Conv} & \multicolumn{1}{c|}{5x5,S1} & 512    \\
\multicolumn{1}{|c|}{PReLu}  & \multicolumn{1}{c|}{-}       & \multicolumn{1}{c|}{-}      & \multicolumn{1}{c|}{PReLu}  & \multicolumn{1}{c|}{-}       & -      \\
\multicolumn{1}{|c|}{S-Conv} & \multicolumn{1}{c|}{5x5,S1} & \multicolumn{1}{c|}{256}    & \multicolumn{1}{c|}{S-Conv} & \multicolumn{1}{c|}{5x5,S1} & 512    \\ \cline{1-3}
\multicolumn{3}{|c|}{\textbf{}}                                                           & \multicolumn{1}{c|}{S-Conv} & \multicolumn{1}{c|}{3x3,S2} & 512    \\ \hline
\multicolumn{6}{|c|}{\textbf{Rate Allocaion Network}}                                                                                                           \\ \hline
\multicolumn{1}{|c|}{Type}   & \multicolumn{1}{c|}{Para.}   & \multicolumn{1}{c|}{Act.}   & \multicolumn{1}{c|}{Out\_c} &                              &        \\ \hline
\multicolumn{1}{|c|}{S-Conv} & \multicolumn{1}{c|}{{\color{blue}3x3,S2}} & \multicolumn{1}{c|}{PReLu}  & \multicolumn{1}{c|}{{\color{blue}256}}    &                              &        \\
\multicolumn{1}{|c|}{S-Conv} & \multicolumn{1}{c|}{{\color{blue}3x3,S1}} & \multicolumn{1}{c|}{PReLu}  & \multicolumn{1}{c|}{{\color{blue}256}}    &                              &        \\
\multicolumn{1}{|c|}{S-Conv} & \multicolumn{1}{c|}{{\color{blue}3x3,S1}} & \multicolumn{1}{c|}{PReLu}  & \multicolumn{1}{c|}{{\color{blue}256}}    &                              &        \\
\multicolumn{1}{|c|}{S-Conv} & \multicolumn{1}{c|}{{\color{blue}3x3,S1}} & \multicolumn{1}{c|}{PReLu}  & \multicolumn{1}{c|}{{\color{blue}256}}    &                              &        \\
\multicolumn{1}{|c|}{{\color{blue}S-Conv}} & \multicolumn{1}{c|}{{\color{blue}3x3,S2}} & \multicolumn{1}{c|}{{\color{blue}PReLu}}  & \multicolumn{1}{c|}{{\color{blue}128}}    &                              &        \\

\multicolumn{1}{|c|}{S-Conv} & \multicolumn{1}{c|}{{\color{blue}3x3,S1}} & \multicolumn{1}{c|}{Sigmoid}      & \multicolumn{1}{c|}{1}       &                              &        \\ \hline
\end{tabular}
\end{table}
The semantic communication framework for perceptual-friendly wireless image transmission is shown in Fig. \ref{fig:image}. As shown in the figure, a semantic encoder is first used to extract semantic information from input images $x$. And then, the VL-SCC module described in Section \ref{spatial} will be adopted to transmit the semantic information from the server side to user side. Finally, a semantic decoder reconstructs the image from recovered semantic information. Denote the reconstructed image as $y$. We show the networks used in this task in Table. \ref{tab:my-table}, where networks that have symmetrical architectures with the listed ones are omitted. {\color{blue} In Table. \ref{tab:my-table}, S-Conv denotes 
signal convolution\footnote{\url{https://www.tensorflow.org/api_docs/python/tfc/layers/SignalConv2D}}, out\_c denotes output channels, and $k\times k$ S1 or S2 denotes $k\times k$ convolution with stride 1 or 2}. In this image transmission task, the training loss is also composed of three terms, a data fidelity term, a semantic loss, and a rate loss.

\textbf{Data fidelity term:}
The data fidelity term is defined as $\textit{l}_{2}$ loss between original images $x$ and reconstructed images $y$. By minimizing this term, we are optimizing pixel-level error.

\textbf{Semantic loss:}
In this task, We use LPIPS loss, $\mathcal{L}_{LPIPS}$, as the semantic loss. LPIPS is defined as the distance of two images in the feature space of a network, defined as $d(x,y)=\sum_{l}\frac{1}{H_{l}W_{l}}\sum_{h,w}||w_{l}\cdot ((F_{x})^{l}_{hw}-(F_{y})^{l}_{hw})||^2_{2}$, where $F^{l}_{x} \in \mathcal{R}^{H_{l}\times W_{l}}$ is the features extracted from $x$ in the $l$-th layer of the network; $w_{l}$ is a weight designed for the $l$-th layer.

\textbf{Rate loss:} The rate loss $\mathcal{L}_{r}$ for the VL-SCC module is defined in Section \ref{spatial}.  

Finally, the total training loss can be represented as,
\begin{equation}
\mathcal{L}=\textit{l}_{2}(x,y)+\lambda \mathcal{L}_{LPIPS}+ \gamma \mathcal{L}_{r},
\label{equ:6}
\end{equation}
where $\lambda$ is an introduced trade-off parameter between data fidelity term and semantic loss.

\section{Experiments}
In this section, we compare semantic communications with traditional communications in human mesh recovery task and wireless image transmission task. We also present experiments to verify the proposed variable-length coding scheme.

\subsection{Human mesh recovery}
In this subsection, we conduct experiments on human mesh recovery task.
\subsubsection{Dataset} We use the following image datasets annotated with 2D joints: LSP, LSP-extended \cite{johnson2010clustered}, and MPII \cite{andriluka20142d}. The adopted image dataset with 3D joints we use is MPI-INF-3DHP \cite{mehta2017vnect}. Due to the lack of 3D joints annotations, we only randomly select 5\% samples for validation, 5\% samples for testing, and use the rest for training. All images are scaled to $224\times 224$ before feeding into ResNet50v2. The images are also randomly scaled, translated, and flipped as a data augmentation method. 

\subsubsection{Experimental environments} We consider additive white Gaussian noise (AWGN) channels, and the signal-to-noise-ratio (SNR) per symbol changes from $-20dB$ to $10dB$. We evaluate the performance of different methods under the same average number of modulated symbols sent by XR users. In this experiment, we set it as $2,000$ ($1,000$ in complex number).

\subsubsection{Considered transmission methods} We now describe the considered transmission methods for human mesh recovery task.
\begin{itemize}
    \item \textbf{benchmark}: In traditional communications, XR users will first transmit the images to the remote server, and the remote server will use received images for task execution. To simulate this process, we first transmit all the training images over noisy channels, and then train a human mesh recovery network based on the received images and true labels. During the image transmission process, we use better portable graphics (BPG) \cite{sullivan2012overview} for source coding, and low-density parity-check (LDPC) for channel coding. For each SNR situation, we first fine-tune the channel coding rate and modulation order to ensure bit-error rate smaller than a threshold. After that, we adjust the BPG compression rate under the bandwidth constraint.
    \item \textbf{VL-SCC}: In this method, we adopt the semantic communication framework designed in Section \ref{sec:human mesh}. The semantic coding module and the VL-SCC module are jointly trained for each SNR situation. We set $N=4,000$, which means the maximum number of modulated symbols available for each sample is $4,000$. By fine-tuning to the value of $\gamma$ in different SNR situations, we ensure that the average number of modulated symbols sent by users is smaller than $2,000$ to satisfy system requirement.
    \item \textbf{SCC(2k)}: By deleting the rate allocation network and its output, we can get a fixed-length semantic communication network, where the $N=2,000$ modulated symbols are transmitted directly to the receiver without any rate control.
    \item \textbf{SCC(4k)}: In this case, we show the performance of fixed-length SCC systems with $N=4,000$, which is unrealistic in the current experimental settings but served as a performance indicator.
\end{itemize}

%\footnote{For each SNR situation, we first fine-tune the channel coding rate and modulation order to ensure bit-error rate smaller than a threshold. After that, we adjust the BPG compression rate under the bandwidth constraint}. 

\subsubsection{Performance metrics}
\begin{figure}[t]
\centering
\includegraphics[scale=0.75]{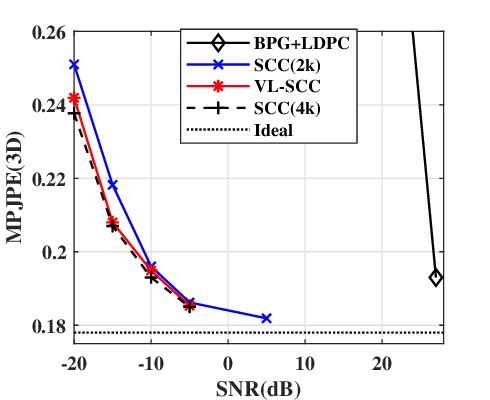}
\caption{MPJPE vs SNR for different transmission methods.}
\label{fig:hmr}
\end{figure}
We use mean per joint position error (MPJPE)\footnote{We do not multiply the predicted joints with the image size here, therefore the reported MPJPE belong to (0,1)} \cite{kanazawa2018end} as the performance matrix, which is calculated based on the true 3D joints labels, $l_{3D}^{T}\in \mathcal{R}^{K\times 3}$, and the estimated 3D joints locations, $l_{3D}^{F}\in \mathcal{R}^{K\times 3}$, where $K$ is the number of joints. Suppose there are $M$ training samples. MPJPE is calculated by $\sum_{i=1}^{M}\sum_{j=1}^{K}\frac{1}{MK}(||{l_{3D}}_{i}^{T}(j)-{l_{3D}}_{i}^{F}(j)||_{2})$. A lower MPJPE indicates the estimated 3D joints locations are more accurate. We show the performance of different transmission methods in Fig. \ref{fig:hmr}, where the ideal case is the performance limit when the source is transmitted over noiseless channels. As we can see from Fig. \ref{fig:hmr}, our proposed designs under different settings work significantly better than the traditional method with BPG and LDPC. This is because we only transmit the task-related information and the information can be protected better when transmitted alone compared with when transmitted alongwith images, as all wireless resources are only used for these task-related information. Moreover, from Fig. \ref{fig:hmr}, VL-SCC outperforms SCC (2K), which verifies the effectiveness of VL-SCC in improving coding efficiency.  

\subsubsection{Semantic loss validation}
\begin{figure*}[t]
\centering
\includegraphics[scale=0.45]{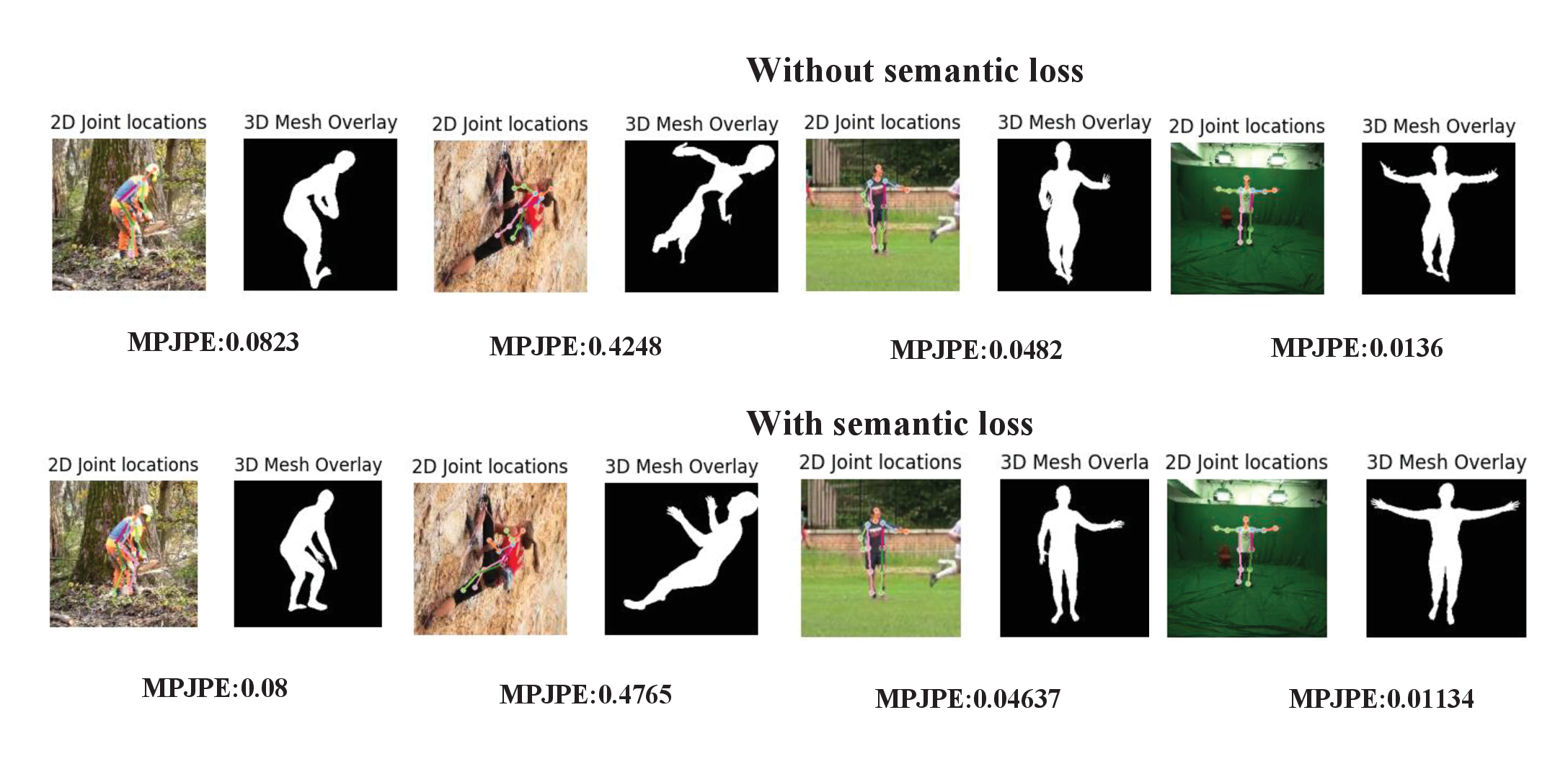}
\caption{The visual effects of human mesh recovery task with/without semantic loss.}
\label{fig:visual}
\end{figure*}
We show the visual effects of human mesh recovery task with and without semantic loss on some testing examples in Fig. \ref{fig:visual}. From the figure, when semantic loss is not considered, the reconstructed human meshes have terrible visual effects, even if the corresponding MPJPE is nearly the same as or even better than the method with semantic loss. This verifies the importance of introducing a semantic loss.
%As discussed above, semantic loss is introduced to ensure the reconstructed human mesh has a good visual quality. To verify the functionality of semantic loss, we show the reconstructed meshes from networks trained with and without semantic loss in Fig..
\subsection{Perceptual-friendly wireless image transmission}
In this part, we consider the wireless image transmission task. The details are as follows.

\subsubsection{Dataset}
 We choose MS-COCO 2014 \cite{lin2014microsoft} as the dataset\footnote{As discussed above, the spatial size of the modulated symbols will be only $1/256$ of the input images to reduce the overhead of the side link. This means the training images cannot be very small. Also, the number of training data should be large enough to show diversity in the amount of semantic information. Based on this considerations, we choose MS-COCO 2014}, in which we use $82,783$ training samples, $2,000$ validation samples, and $2,000$ testing samples. During the training, $128\times 128$ image patches are randomly cropped.

\subsubsection{Experimental environments} We consider additive white Gaussian noise (AWGN) channels, and the signal-to-noise-ratio (SNR) per symbol is set as $10dB$. The average number of modulated symbols sent by XR servers varies in this experiment.

\subsubsection{Considered transmission methods} We now describe the considered transmission methods for semantic-aware wireless image transmission.
\begin{itemize}
    \item \textbf{BPG+LDPC:} Similar to the previous experiment, we use BPG for source coding and LDPC for channel coding. {\color{blue} Following the experimental settings in \cite{dai2022nonlinear}, when SNR$=10dB$, we use 16QAM and $2/3$ LDPC}.
    \item \textbf{VL-SCC($\lambda$):} In this method, we adopt the semantic communication framework design in Section \ref{sec:image}. We set $N=512$, which is the maximum code length for each spatial location. We fine-tune the value of $\gamma$ to adjust average number of modulated symbols. When $\lambda=0$, the network is not trained with LPISP loss.
    \item \textbf{SCC($\lambda$):} This is the fixed-length coding version of VL-SCC by neglecting the rate allocation network. We adjust the average number of modulated symbols by changing the channel dimension of the modulated symbols.
    \item \textbf{JSCC:} JSCC proposed in \cite{bourtsoulatze2019deep} is also presented as a benchmark of deep learning based image transmission scheme. {\color{blue} We use the JSCC architecture proposed in \cite{kurka2020deepjscc} for this method.}
\end{itemize}

\subsubsection{Performance metrics}
\begin{figure}[t]
\centering
\includegraphics[scale=0.75]{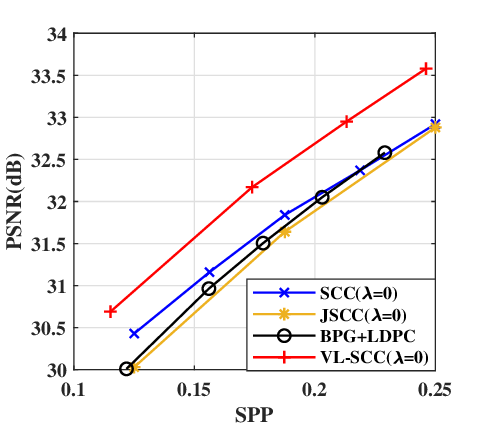}
\caption{{\color{blue}PSNR vs symbols per pixel for different transmission methods.}}
\label{fig:PSNR}
\end{figure}
We first compare the peak signal-to-noise ratio (PSNR) versus the number of symbols per pixel (SPP) of different transmission methods, where the SPP is defined as the average number of modulated symbols (in complex numbers) assigned to each color pixel. The PSNR value reflects the per-pixel reconstruction quality of images. A higher value indicates a better reconstruction quality. The results are shown in Fig. \ref{fig:PSNR}. From the figure, our fixed-length coding method SCC($\lambda=0$) outperforms the JSCC and traidtional methods with BPG and LDPC as the source coding and channel coding schemes, respectively, showing the semantic coders and the SCC module used in our work are effective. Also, the proposed variable-length coding method VL-SCC($\lambda=0$) can {\color{blue}significantly} increase the coding efficiency {\color{blue}of SCC}, which also shows the proposed VL-SCC is universal and can be applied to various tasks and data types.
\begin{figure}[t]
\centering
\includegraphics[scale=0.75]{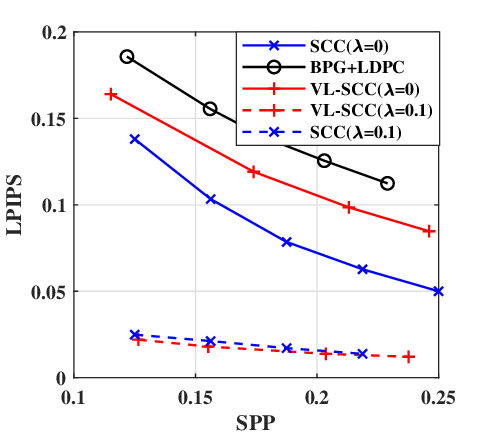}
\caption{{\color{blue}LPIPS vs symbols per pixel for different transmission methods.}}
\label{fig:lpips}
\end{figure}

The LPIPS loss versus SPP is shown in Fig. \ref{fig:lpips}. From the figure, without giving a LPIPS loss during training, VL-SCC($\lambda=0$), SCC($\lambda=0$), and BPG all lead to high semantic errors. However, once the semantic error is considered during the training processing, such as VL-SCC($\lambda=0.1$) and SCC($\lambda=0.1$), it can be suppressed  significantly. {\color{blue}Furthermore, the proposed variable-length coding method VL-SCC($\lambda=0.1$) also outperforms the fixed-length method SCC($\lambda=0.1$ in semantic measurements.}

\section{Conclusion}
\label{sec:conclusion}
In this work, we design a semantic communication framework for XR. Particularly, we develop the semantic coding modules for different XR tasks, including light estimation, 3D mesh recovery, image transmission, and so on. We propose a novel differentiable variable-length coding mechanism for semantic communications, where the best code length allocation scheme is learnt in an end-to-end manner. Moreover, we verify the effectiveness of the proposed semantic coding scheme and variable-length semantic-channel coding scheme in both the uplink and downlink of wireless XR, showing the generality of the proposed methods.    
\bibliographystyle{IEEEtran}
\bibliography{ref}
\small
\end{document}